\documentstyle[12pt,epsfig]{article}     
\begin{document}

\title{Reinterpreting several narrow `resonances' as threshold cusps}

\vskip 3mm
\begin {center}
{D.V.~Bugg},   \\
{Queen Mary, University of London, London E1\,4NS, UK}
\end {center}
\vskip 2.5mm

\begin{abstract}
The threshold $\bar pp$ peak in BES data for $J/\Psi \to \gamma \bar
pp$ may be fitted as a cusp.
It arises from the well known threshold peak in $\bar pp$
elastic scattering due to annihilation.
Several similar examples are discussed.
The PS185 data for $\bar pp \to \bar {\Lambda }\Lambda$ require an
almost identical cusp at the $\bar {\Lambda }\Lambda$ threshold.
There is likewise a cusp at the $\Sigma N$ threshold in
$K^-d \to \pi ^- (\Lambda p)$.
Similar  cusps are likely to  arise at thresholds for all 2-body
de-excitation processes, providing the interaction is attractive;
likely examples are $\Lambda \bar p$, $\Sigma  \bar p$,
and $\bar K \Lambda$.
The narrow peak observed by  Belle at 3872 MeV in
$\pi ^+ \pi ^- J/\Psi$ may be a $J^{PC}=1^{++}$ cusp due to the
$D\bar D^*$ threshold.
The narrow $\Xi ^*(1862)$ observed by NA49
may be due to a threshold cusp in $\Sigma (1385)\bar K $ coupled to $\Xi
\pi$ and $\Sigma \bar K$.
The relation of cusps to known resonances such as $f_0(980)$ is
discussed.
\end{abstract}

Wigner pointed out that a cusp appears in the cross section for any
process at the threshold where a coupled channel opens [1].
Such cusps were studied in the 1958-61 era by Baz' and Okun [2],
Nauenberg and Pais [3] and others.
T\" ornqvist has emphasised the importance of cusps in meson-meson
scattering and their relation to resonances [4].

Several narrow peaks attributed to resonances
may in fact be cusp effects.
The cusp has a different structure to a resonance: the behaviour
of the real part of the amplitude is quite different.
Under some circumstances, a threshold can induce or capture a
resonance.
The conditions under which a resonance is likely to be trapped are
discussed using $f_0(980)$ as an example.

The BES collaboration reports a threshold $\bar pp$ peak in
$J/\Psi \to \gamma \bar pp$ [5], and fits it as a narrow resonance
just below the $\bar pp$ threshold. Datta and O'Donnell conjecture a narrow
quasi-bound state of $\bar pp$ [6].
The Belle collaboration has also reported low mass $\bar pp$
peaks in $B^+ \to K^+\bar pp$ [7] and
$\bar B^0 \to D^0 \bar pp$ [8].

In view of the very large number of open channels known in
$\bar pp$ annihilation at rest, a narrow resonance is surprising.
Why should such a resonance be narrow compared to conventional
meson widths of $\sim 250$ MeV?

There are threshold peaks in the $\bar pp$ total cross
section [9] and annihilation cross section [10,11]; both rise
continuously towards threshold and may be parametrised as $A + B/k$,
where $k$ is centre of mass momentum.
The $B/k$ term follows the familiar `$1/v$ law' of thermal
neutron physics and is symptomatic of absorption from the
$\bar pp$ channel into other open channels [12].

The process $J/\Psi \to \gamma \bar pp$ will be discussed following
Watson's treatment of final-state interactions [13]. The production
process is considered in terms of two vertices.
The first produces $\gamma X$; most details
of the production mechanism will be neglected, therefore
absolute cross sections cannot be predicted. Attention
will be focusssed on a second vertex where all channels having the
quantum numbers of $X$ participate in a final-state interaction. From
this second vertex, the $\bar pp$ channel is one of the emergent
channels.
If the second vertex is resonant, one arrives at
the conventional Isobar Model.
Rescattering between the spectator photon and decay products of $X$ is
neglected.

The final-state interaction may also  be non-resonant:
a familiar example is $\pi ^- d \to \gamma (nn)$, which provides
one of the best measurements of the $nn$ scattering length.
Suppose the amplitude $f_S$ for $\bar pp$ elastic scattering is
written in the $N/D$ form, where $N(s)$ has only left hand
cuts and $D(s)$ has only right-hand cuts.
The content of Watson's theorem is that $D(s)$ is the
same for all channels in which $\bar pp$ appears.
The structure in BES data should be the same as in $\bar pp$
elastic scattering.

The data will be fitted using a scattering length approximation
$k\cot \delta = 1/a$, where $a$ is complex.
If the S-wave amplitude is written as $f_S = (e^{2i\delta } - 1)/2ik$,
simple K-matrix algebra gives
\begin {eqnarray}
f_S &=& \frac {a}{1 - iak} =
\frac {a + i|a|^2k}{1 + 2k\, \mathrm {Im}\, a + k^2|a|^2}, \\
|f_S|^2 &=& \frac {|a|^2}{1 + 2k\, \mathrm {Im}\, a +k^2|a|^2}.
\end {eqnarray}
Eqn. (2) expresses the enhancement factor for a non-resonant
final-state interaction.
The $k$-dependent terms are due  to unitarity and guarantee
that $f_S$ obeys the unitarity limit for large $k$.
The total cross section $\sigma _{tot} = 4\pi \mathrm{Im}\,
f_S/k$,
follows the $1/v$ law, as does the inelastic cross section.

\begin{figure}
\begin{center}
\epsfig{file=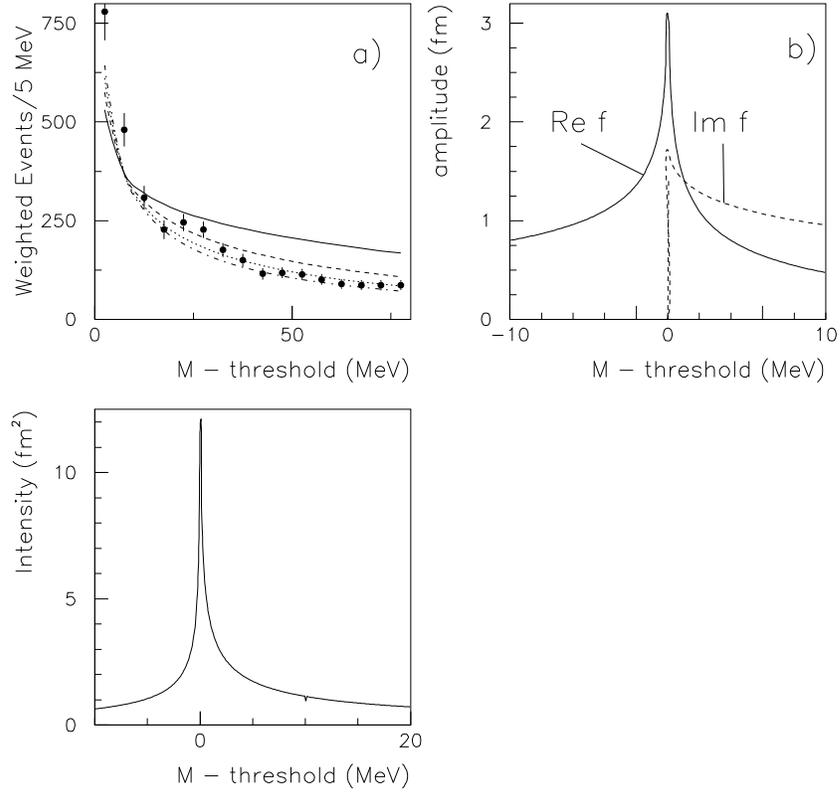,width=12cm}\
\caption{(a) The $\bar pp$ mass spectrum observed by BES, after
dividing out S-wave phase space.
Curves show fits to a scattering length approximation with
$\mathrm {Im}\, a = 0.6$ fm (full curve), 1.2 fm (dashed),
1.8 fm (dotted) and 2.4 fm (chain), including the dispersive
corrections to $\mathrm {Re}\, f_S$;
(b) real and imaginary parts of $f_S$ within 10 MeV of threshold;
(c) $|f_S|^2$, including its analytic continuation below threshold.}
\end{center}
\end{figure}

There is a step in $\mathrm {Im}\, f_S$ at threshold (since it vanishes
below threshold). The real part of the amplitude is given by a
dispersion relation:
\begin {equation}
\mathrm {Re}\, f_S(s) = \frac {1}{\pi }\, P \int \frac
{\mathrm {Im}\, f_S(s')\, ds'}{s' - s}.
\end {equation}
If  $\mathrm {Im}\, f_S$ were strictly constant, $\mathrm {Re}\, f_S$
would be logarithmically divergent:
\begin {equation} \mathrm {Re}\,
f_S =  \frac {\mathrm {Im}\, a}{\pi } \ln
\left( \frac {4M^2 - s_0}{|4M^2 - s|} \right).
\end {equation}
The full eqns. (1) and (3) give a convergent but
similar peak in $\mathrm {Re}\, f_S$ at threshold, as illustrated
in Fig. 1(b).
The peak is positive both below and above threshold.
The dispersive peak in $\mathrm{Re}\, f_S$ represents an effective
attraction, which can lock a resonance at or close to the threshold.

Fig. 1(a) shows BES data, after dividing out the S-wave phase
space factor, as in their Fig. 3(b), from which the data points
are taken.
Curves show results from
eqn. (2) for several values of $\mathrm {Im}\, a$.
The real part of the amplitude is included from eqn. (3) using a
subtraction at $k = 110$ MeV/c, where Coulomb interference data
give $\rho = \mathrm {Re}\, f(0)/\mathrm {Im}\,
f(0)\sim 0$ [14].
The best fit requires
$\mathrm {Im}\, a \simeq 1.8$ fm (dashed curve).
Fig. 1(b) is plotted with this value. However, this number is not
well determined because
(i) there could well be an effective range term in $f_S$, (ii) there is
the possibility of some P-wave contributions to BES data at the higher
masses.

Production of the $\bar pp$ $^1S_0$ final state in $J/\Psi \to
\gamma \bar pp$ requires orbital angular momentum $L=1$ at the
production vertex.
A factor $E^3_\gamma$ is included into the
production cross section, since the $c\bar c$ interaction is
`pointlike'; this factor enhances the lowest $\bar pp$ masses slightly.
The cross section is also enhanced by the Coulomb
attraction near threshold by a factor $(1 - e^{-X})/X$, where
$X = \pi\alpha/\beta$ and $\beta ^2 = 1 - 4M^2/s$ [15]; $M$ is
nucleon mass.
This factor affects only the first two points significantly.
The lowest point is enhanced by $19\%$ and the highest by 4.4\%, so
Coulomb attraction does not account for the peak.


The $\bar pp$ annihilation cross section is close to the unitarity limit
[16].
The cross section for $\bar pp \to \bar nn$ is very small.
The amplitude for this process depends on $f_S(I=1) - f_S(I=0)$
and therefore
requires an accurate cancellation between the imaginary parts of
amplitudes for the two isospins. If the threshold peak were due to a
narrow resonance, this would require two $I=1$ and $I=0$ resonances
accurately degenerate in mass, width and coupling strength. Such a
triple coincidence is implausible, since attractive forces from meson
exchanges are likely to be significantly different for the two
isospins. This is a first argument against a resonance interpretation.

\begin{figure}
\begin{center}
\epsfig{file=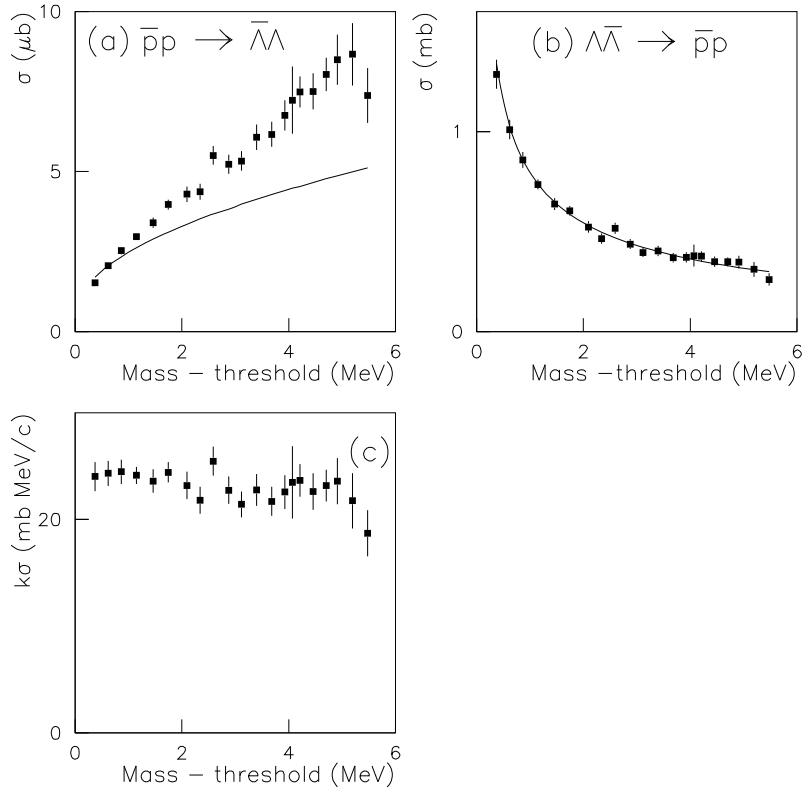,width=12cm}\
\caption{(a) PS 185 integrated cross sections for $\bar pp \to \bar
\Lambda \Lambda$; the curve shows the S-wave cross
section from a phase shift analysis; (b) the corresponding
cross section for $\bar \Lambda \Lambda \to \bar pp$, after
subtracting P-wave cross sections; the curve is a fit using
eqns. (2) and (3);
(c) $\sigma (\bar \Lambda \Lambda \to \bar pp ) \times k $ v.
excitation energy after subtracting P-waves.}
\end{center}
\end{figure}
Consider next $\bar pp \to \bar {\Lambda }\Lambda$.
The PS185 collaboration has measured cross sections
at fine steps of $\bar p$ momentum very close to threshold [17].
A full partial wave analysis, including extensive spin dependent data,
is reported elsewhere [18].
The data points of Fig. 2(a) show integrated cross sections.
The curve shows the cross section fitted to the $\bar \Lambda
\Lambda$ S-wave;
the difference from data is due  to $\bar \Lambda
\Lambda $ P-waves, which are well determined by polarisation data
and the forward-backward asymmetry in $d\sigma /d\Omega$.
The  P-waves are surprisingly large.
A possible reason, proposed by PS185, is that P-waves are
highly peripheral, and do not suffer from absorption into other
channels; in the mass range shown in Fig. 2, $k < 85$ MeV/c
corresponding to an impact parameter $>2$ fm.

Using detailed balance, the cross section for the inverse process
may be derived:
\begin {equation}
\sigma (\bar {\Lambda }\Lambda \to \bar pp) = \frac {k^2_p}{k^2_\Lambda
} \sigma (\bar pp \to \bar {\Lambda }\Lambda ),
\end {equation}
where  $k_p$ and $k_\Lambda$ are centre of mass momenta of $p$ and $\bar
\Lambda$.
Fig. 2(b) shows resulting cross sections after subtracting
P-wave contributions.
There is a definite  cusp at the $\bar {\Lambda }\Lambda$ threshold.
It has not been reported before.
Fig. 2(c) shows that this cross section fits a $1/k$ dependence.
The peak of Fig. 2(b) appears narrower than the $\bar pp$ peak of
Fig. 1(a) because one is looking at  different parts of the $1/v$
curve, but they can both be fitted with an identical
value of $\mathrm {Im}\, a$.
The cross section of Fig. 2(b) reaches 1 mb at 0.5 MeV excitation
energy.
If the $\bar \Lambda \Lambda$ annihilation cross section
reaches its unitarity limit ($> 500$ mb at the same energy),
$\bar \Lambda \Lambda \to \bar pp$ can only be one of a large
number of open channels.

A third example of a cusp is in $K^-d \to \pi ^- \Lambda p$,
where a peak is observed [19,20] in the $\Lambda p$ mass spectrum
at the $\Sigma N$ threshold.
My fit with a cusp is shown in Fig. 3, using the weighted mean
of data  from Refs. [19] and [20].
Separate cusps are fitted to $\Sigma ^0p$ and $\Sigma ^+n$, weighted
in the ratio 2: 1 of Clebsch-Gordan coefficients for
$I = {1 \over 2}~ (\Lambda p)$.
The optimum fit requires shifting experimental data upwards by 1 MeV,
within their errors.

\begin{figure}
\begin{center}
\epsfig{file=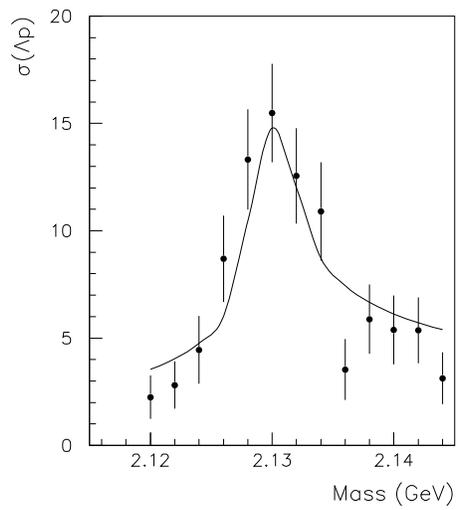,width=8cm}\
\caption{The $\Lambda p$ mass spectrum from the weighted mean of
data from Refs. [18] and [19]; the curve shows the fit to a
cusp. The vertical scale is unnormalised.}
\end{center}
\end{figure}

The data have also been fitted by Dosch and Stamatescu [21], who
conclude that there is a $^3S_1$ pole.
Earlier, Nagels, Rijken and de Swart [22] fitted hyperon-nucleon
scattering data at low energies and required a $\Sigma ^+n$ pole at
2131.77 - i2.39 MeV, very close to the
$\Sigma n$ threshold.
Its width is so small that it is difficult to
separate from a cusp effect with existing data.

Next, the Belle collaboration has presented evidence for a very
narrow $\pi ^+\pi ^-J/\Psi$ peak at 3872 MeV in $B^\pm \to K^\pm (\pi
^+\pi ^-J/\Psi )$ [23].
This mass is 0.8 MeV above the $D^0\bar D^{0*}$ threshold and 7 MeV
below that for $D^\pm \bar D^{\mp *}$.
The peak could well be due to an S-wave cusp expected
at the  mean threshold of 3875 MeV.

Suppose $D\bar D^*$ final states are produced randomly in the
production process.
Those close to threshold follow a $1/v$ cross section for
de-excitation to open channels.
[A proviso is that the $\bar DD^*$ interaction is attractive near
threshold; if the real part of the interaction is repulsive, the
wave function at low momentum may be shielded from short range
annihilation.]
There are many open channels: $J/\Psi \rho $, $[\eta _c(\pi
\pi)_S]_ {L=1}$, $[\chi _{c0}\pi ]_{L=1}$, $[\chi _{c1}\pi ]_{L=1}$ and
$\chi _{c1}(\pi \pi )_S$, where $(\pi \pi )_S$ denotes the $\pi \pi$
S-wave.
The observed cross section in a final state such as $J/\Psi \rho$
will be given by $D\bar D^*$ phase space multiplied into
the de-excitation cross section.
The $1/k$ dependence of the cross section is almost cancelled by
the $k/\sqrt {s}$ dependence of the phase space;
the variation of $J/\Psi \rho$ phase space over the narrow region
being considered is also negligible.
Neglecting these two factors, the result is given by eqn. (2).

Fig. 1(c) illustrates the corresponding result for any channel fed by
$p\bar p$.
The peak comes from the dispersive effect in the real part
at threshold.
If the final state is fed entirely by $\bar pp$ annihilation,
the peak  will be cut off sharply below threshold.
However, it is also possible for the final
states to be produced via other mechanisms. These will have
precisely the same $s$-dependence, since the final state
`knows' about the threshold through analyticity.
The dispersive peak marks the opening of the 2-body channel,
both in $\bar pp$ and in $D\bar D^*$.

Further threshold cusps may arise in all annihilation
channels involving narrow particles, providing
the interaction is attractive, so that the annihilation is not
suppressed.
Possible examples are $D\bar
D$ (due to decay to $J/\Psi \rho$ and $\chi _{C0}(\pi \pi )_S$),
$D^*\bar D^*$ ($J^P=2^+$, $1^+$ or $0^+$ with many open channels),
$\Xi \bar p$, $\Sigma \bar p$, $\Lambda \bar p$ and so on.


Example 5 is the narrow peak at 1862 MeV observed by the NA49
collaboration in
$\Xi ^- \pi ^-$, $\Xi ^-\pi ^+$ and their charge conjugates [24].
It requires exotic quantum numbers $I=3/2$ and has been proposed
as a pentaquark.
The $\Sigma (1385)\bar K$ threshold lies slightly higher.
The $\Sigma (1385)\bar K$ pair can de-excite to $\Sigma \bar K$,
$\Xi \pi$ and $\Xi ^*(1530)\pi$.
An illustration of the process is shown in Fig. 4(a).
All these processes are likely to have $1/v$ cross sections
near the $\Sigma (1385)\bar K$ threshold.

\begin{figure}
\begin{center}
\epsfig{file=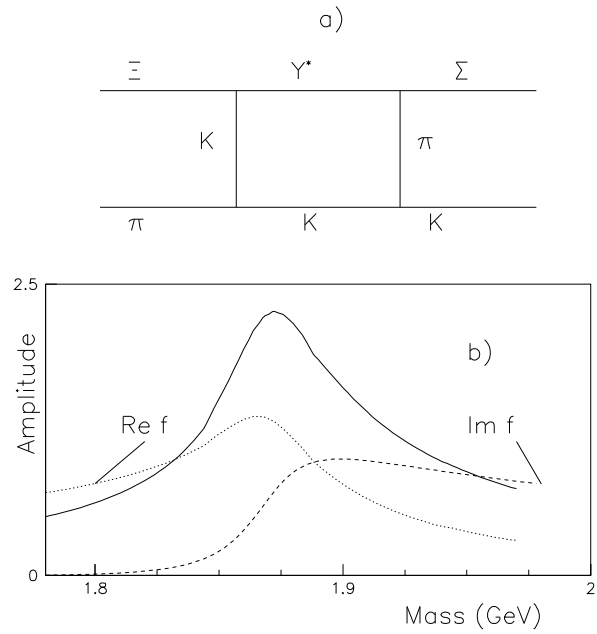,width=10cm}\
\caption{(a) The Graph for de-excitation of $\Sigma (1385)\bar K$
to $\Xi \pi$ and $\Sigma K$; (b) Real (dotted) and Imaginary (dashed)
parts of the $\Sigma (1385)\bar K$ elastic amplitude and its
intensity (full curve).}
\end{center} \end{figure}

Using eqn. (2) , it is a simple matter
to fold the energy dependence of the de-excitation process
$\Sigma (1385)\bar K \to \Xi \pi$ with the line-shape of
$\Sigma (1385)$.
The calculation assumes a $\Sigma (1385)\bar K$
scattering length of 1.8 fm, as for $\bar pp$; however, there is
very little sensitivity to this value, except for the absolute
normalisation.
The calculation uses a P-wave line-shape of
$\Sigma (1385)$ with a centrifugal barrier radius of 0.8 fm.
It also includes mass differences between charge states with a
weighting for charges  taken from the NA49 data.
Fig. 4(b) shows Real and
Imaginary parts of the $\Sigma (1385)\bar K$ elastic scattering
amplitude.
The imaginary part rises like a Fermi function due to the width
of $\Sigma (1385)$.
The intensity is shown in Fig. 4(b) by the full curve. It gives
rise to a peak in de-excitation channels, e.g. $\Xi ^- \pi ^-$, centred
at 1872 MeV. The NA49 collaboration quotes a
mass of $1862 \pm 2$ MeV for $\Xi ^- \pi ^-$ and $1864 \pm 5$ MeV for
$\Xi ^-\pi ^+$. So there remains a discrepancy of $\sim 9 \pm 3$ MeV
with the predicted mass. This discrepancy could arise from interference
with background amplitudes. The width quote by NA49 also appears
to be smaller: $< 18$ MeV.

There is a useful and well documented analogue in heavy ion
elastic scattering at the Coulomb barrier,
observed in a wide variety of examples throughout the
nuclear periodic table.
The topic is reviewed by Satchler [25].
Real and imaginary parts of the optical potential are derived
from accurate experimental data.
The imaginary part of the potential rises swiftly as the Coulomb
barrier is overcome and inelastic channels open.
This is analogous to the dashed curve of Fig. 4(c).
The real part of the potential peaks at the centre of the
leading edge, like the peak in $\mathrm {Re}\, f_S$.
Satchler's Fig. 2.2 for $^{16}O + ^{208}Pb$ elastic scattering is
very similar to Fig. 4(b).

A final example of a cusp is in  $\pi d \to NN$ below 10 MeV [26].
In this case, it is well known that there is no $^3P_1$ $NN$
resonance at this threshold.
In fact, the $NN$ $^3P_1$ phase shift is repulsive.

In all these cases, cusps may account for the data,
but there is also the possibility of a resonance interpretation.
To understand whether a resonance is likely, it is
instructive to consider $f_0(980)$ as an example.
This resonance is fitted with the Flatt\' e form:
\begin {eqnarray}
f_S &=& 1/[M^2 - s - m(s) - iM(g^2_\pi\rho _{\pi \pi}(s) +
g^2_K\rho _{K\bar K}(s))],  \\
m(s) &=& \frac {M^2 - s}{\pi }\int \frac {M\Gamma _{tot}(s')\, ds'}
{(M^2 - s')(s' - s)},
\end {eqnarray}
where $g$ are coupling constants and $\rho$ is 2-body phase space
$2k/\sqrt {s}$.
Parameters will be taken from the
latest BES data on $J/\Psi \to \phi \pi ^+\pi ^-$ and $\phi K^+K^-$,
where the $f_0(980)$ is particularly well determined in both
$\pi \pi$ and $KK$ decay modes [27].

\begin{figure}
\begin{center}
\epsfig{file=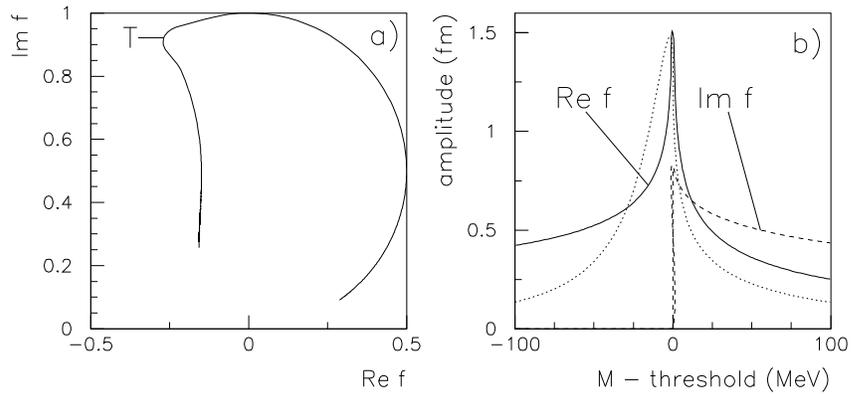,width=12cm}\
\caption{(a) The Argand diagram for $f_0(980)$; $T$ marks the
$K\bar K$ threshold; (b) $\mathrm {Re}\, f$ (full curve),
$\mathrm {Im}\, f$ (dashed) from eqns. (2) and (3) compared with
the actual line-shape of $f_0(980)$ (dotted).
}
\end{center}
\end{figure}

The Argand diagram for the $\pi \pi \to \pi \pi$ amplitude is shown on
Fig. 5(a).
The $K\bar K$ threshold opens at the point $T$,
creating a cusp.
The amplitude is
\begin {eqnarray} \nonumber
f(\pi \pi \to \pi \pi)
&=& \frac {1}{k_\pi}
\frac {2g^2_\pi  k_\pi/\sqrt {s}}
{M^2 - s - m(s) - 2iM(g^2_\pi k_\pi  + g^2_K k_K )/\sqrt {s}} \\
&=& \frac {2g_\pi ^2 /\sqrt {s}}
{M^2 - s - m(s) - 2iM(g^2_\pi k_\pi  + g^2_K k_K)/\sqrt {s}}.
\end {eqnarray}
Amplitudes for $K\bar K \to K\bar K$ and $K\bar K \to \pi \pi$
are obtained by replacing $g^2_\pi$ by $g^2_K$ or $g_\pi g_K$;
note that all three amplitudes share the same dependence on $s$, i.e.
the same denominator $D(s)$.

The dispersive contribution to $\mathrm {Re}\, f$
from eqns. (2) and (3) is shown in Fig. 5(b),
taking $\mathrm {Im}\, a = 0.87$ fm from the Flatt\' e fit to data.
It spreads over a wider mass range than for $\bar pp$, Fig. 1(b),
because $\mathrm {Im}\, a$ is smaller.
The dotted curve on Fig. 5(b) shows the actual line-shape of
the Flatt\' e formula.
There is quite a good overlap between the
dispersive component of $\mathrm {Re}\, f_S$ and the line-shape of the
resonance.
One must remember that there are also attractive forces due
to meson exchanges [28,29] and/or attraction at the quark level.
These add coherently to the dispersive contribution to
$\mathrm {Re}\, f_S$ and play a role in deciding whether or not a
resonance appears.
Janssen et al. remark that the attraction arising from the
$K\bar K$ threshold is important in their model of $f_0(980)$.

For a resonance to develop, the attraction must overcome zero-point
energy, which is large if the wave function is tightly constrained
to small $r$.
A segment of the line-shape with binding energy $B$ beneath the
$K\bar K$ threshold has a radial wave function $\propto e^{-\alpha
r}/r$, where $\alpha = 1/\sqrt {M_KB}$ and $M_K$ is the kaon mass.
The smaller the value of $B$, the lower is the zero-point energy.
The part of the wave function outside the short-range
attraction contributes negatively to zero-point energy; the
wave function is exponentially damped rather than oscillatory.

The ideal circumstance for a resonance locked to the threshold arises
when a single channel (denoted 1) above threshold is coupled weakly
to one other channel (2) below threshold, as for $f_0(980)$ and
$a_0(980)$.
If wave function leaks away into many other channels,
(i) the resonance acquires a width $g_i^2\rho _i(s)$ through coupling
    to each such channel $i$,
(ii) $\mathrm {Re}\, f_S(s)$ arising from channel 1 is weakened because
    wave function is lost from that channel.

If one views the $\bar pp$ and $\bar \Lambda \Lambda$
peaks in this light, resonances are unlikely.
For $\Sigma n \to \Lambda p$, the data lie close to the
unitarity limit, so again the cusp interpretation appears more
likely.
Nagels et al. [22] circumvent this by putting their bound state
so close to threshold that most of the wave function
lies far outside the range of nuclear interactions.

The branching fraction measured
for $J/\Psi \to \gamma \bar pp$ is $7 \times 10^{-5}$ [5].
However, there are much larger branching fractions for
$J/\Psi \to \gamma X$ with $X$ having the same quantum numbers
$J^{PC} = 0^{-+}$, $I=0$.
These channels are $\rho \rho$, $\omega \omega$, $K^*\bar K^*$, $\eta
\pi \pi$ and $K\bar K\pi$; their combined branching fraction is $(1.9
\pm 0.3) \times 10^{-2}$ from Fig. 2 of Bugg, Dong and Zou [30].
This is a second argument against a resonance.
The
BES collaboration sees no threshold $\bar pp$ peak in the final state
$\pi ^0 \bar pp$. This final state has a branching fraction $10^{-3}$,
much larger than $\gamma \bar pp$. It is likely to be dominated by
$N^*\bar N$ and $\Delta \bar N$ channels; their angular momenta have
only small overlap (Racah coefficients) with $N\bar N$
S-waves.

In summary, cusps are capable of explaining in a simple way
peaks observed at many thresholds. These cusps are a
direct consequence of decay to open channels.
The cusp is driven by the peaking of the S-wave de-excitation cross
section due to the $1/v$ law. The
singularity at a cusp is of the form $a/(1 - iak)$ and has a real part
different from a resonance. A resonance is to be expected only under
restrictive circumstances such as those for $f_0(980)$, where there is
a single weak open channel.

While this work was being written up, a related paper has appeared
from Kerbikov, Stavinsky and Fedotov [31].
They also attribute the narrow structure in $\bar pp$ to a cusp
and fit it with the scattering length approximation; they do not
however consider the dispersive contribution to $\mathrm {Re}\, f_S$.

I am grateful to Dr. T. Johansson for tables of PS 185 data
and to Prof. J. de Swart for comments on hyperon-nucleon scattering.

\end {document}